\documentclass[journal]{IEEEtran} 

\usepackage{amsmath}
\usepackage{amssymb,amstext}
\usepackage[center]{subfigure}
\usepackage{lineno,hyperref}
\usepackage{xcolor}
\usepackage{extarrows}
\usepackage{algorithm}
\usepackage{algorithmic}
\usepackage{setspace}
\usepackage{bm}
\usepackage{graphicx}
\usepackage[justification=centering]{caption}
\usepackage{cite}
\usepackage{subfigure}
\usepackage{url}
\usepackage{multirow}
\usepackage{tabularx}
\usepackage{booktabs}
\usepackage{tikz}
\usepackage{amsfonts}
\usepackage{float}
\usepackage{array}
\usepackage{cite}
\usepackage{cases} 
\usetikzlibrary{arrows,positioning,fit,shapes,calc,decorations.pathreplacing,matrix,patterns}

\newcommand{\larrow}{\overset{\scriptscriptstyle\leftarrow}}   
\newcommand{\rarrow}{\overset{\scriptscriptstyle\rightarrow}}

\begin{document}
\title{{Hierarchically Structured Matrix Recovery-Based Channel Estimation for  RIS-Aided Communications}} 
	%	Hybrid Message Passing Based Low-Cost Channel Estimation for  RIS-Aided MIMO Communications}}
	%A Low-complexity Channel Estimation for RIS-Aided MIMO Communications} 
	%Based {on} Sparse Bayesian Learning with UAMP}
\author{Yabo Guo, Peng Sun, Zhengdao Yuan, Qinghua Guo, \IEEEmembership {Senior Member, IEEE} and Zhongyong Wang 
    \thanks{The work of Y. Guo, P. Sun, Z. Yuan and Z. Wang was supported by National Natural Science Foundation of China (61901417).
	%(\emph{Corresponding authors: Qinghua Guo and Zhongyong Wang}.)
}
	\thanks{Y. Guo, P. Sun and Z. Wang are with the School of Information Engineering, Zhengzhou University, Zhengzhou 450002, China (e-mail: ieybguo@163.com, iepengsun@zzu.edu.cn,  zywangzzu@gmail.com).}
	\thanks{Z. Yuan is with the Artificial Intelligence Technology Engineering Research Center, Open University of Henan, Zhengzhou 450002, China (e-mail: yuan\_zhengdao@163.com).}
%	\thanks{C. Huang is with the College of Information Science and Electronic Engineering, Zhejiang University,
%	Hangzhou 310007, China, and Zhejiang Provincial Key Lab of Information Processing, Communication and Networking
%	(IPCAN), Hangzhou 310007, China, and the International Joint Innovation Center, Zhejiang University, Haining 314400, China
%	(e-mail: chongwenhuang@zju.edu.cn).}
	\thanks{Q. Guo is with the School of Electrical, Computer and Telecommunications Engineering, University of Wollongong, Wollongong, NSW 2522, Australia  (e-mail: qguo@uow.edu.au).}
%	\thanks{C. Yuen is with the Engineering Product Development (EPD) Pillar, Singapore University of Technology and
%	Design, Singapore 487372 (e-mail: yuenchau@sutd.edu.sg).}
}
\maketitle

\begin{abstract}
Reconfigurable intelligent surface (RIS) has {emerged} as a promising technology for improving capacity and extending coverage {of wireless networks. In this work, we consider RIS-aided millimeter wave (mmWave)  multiple-input and multiple-output (MIMO) communications, where acquiring accurate channel state information is challenging due to the high dimensionality of channels. To fully exploit the structures of the channels, we formulate the channel estimation as a hierarchically structured matrix recovery problem, and design a low-complexity message passing algorithm to solve it. %leveraging unitary approximate message passing. 
Simulation results demonstrate the superiority of the proposed algorithm and its performance close to the oracle bound.}% and the superiority of the proposed estimator with light training overhead and latency.
%Reconfigurable intelligent surface (RIS) is very promising for wireless networks to achieve high energy efficiency, extended coverage, improved capacity, massive connectivity, etc. To unleash the full potentials of RIS-aided communications, acquiring accurate channel state information is crucial, which however is very challenging. 
%For RIS-aided multiple-input and multiple-output (MIMO) communications, the existing channel estimation methods have computational complexity growing rapidly with the number of RIS units $N$ (e.g., in the order of $N^2$ or $N^3$) and/or have special requirements on the matrices involved (e.g., the matrices need to be sparse for algorithm convergence to achieve satisfactory performance), which hinder their applications. 
%In this work, instead of using the conventional signal model in the literature, we derive a new signal model obtained through proper vectorization and reduction operations. Then, leveraging the unitary approximate message passing (UAMP), we develop a more efficient channel estimator that has complexity linear with $N$ and does not have special requirements on the relevant matrices, thanks to the robustness of UAMP. These facilitate the applications of the proposed algorithm to a general RIS-aided MIMO system with a larger $N$. Moreover, extensive numerical results show that the proposed estimator delivers much better performance and/or requires significantly less number of training symbols, thereby leading to notable reductions in both training overhead and latency. 
\end{abstract}


\section{Introduction}
\IEEEPARstart {A}{s} 
a promising technology in future wireless communications, reconfigurable intelligent surface (RIS) is capable of altering the wireless propagation environment to achieve desired channel responses by dynamically adjusting the massive number of passive reflecting elements. With the aid of RIS, wireless networks can achieve high energy efficiency, improve the system capacity and coverage, {and} enhance massive connectivity \cite{WangSPL2020,Huang2019TWC,PengTCOM2022,HeWCL2020,LiuJSAC2020,WangTWC2020,ZhangCL2022,GuoTWC2023}. 

In order to {realize} the potential of RIS-aided communications, {the acquisition of accurate channel state information (CSI) is crucial, which however is very challenging especially in RIS-aided multiple input multiple output (MIMO) systems due to the high dimensionality of channels.}
{The work in \cite{PengTCOM2022} studied the uplink channel estimation protocol for a RIS-aided mmWave system, which assumes {a} random Bernoulli RIS phase matrix.} The work in \cite{HeWCL2020} formulated the downlink block fading channel estimation as a sparse matrix factorization and completion problem, {where the low-rank or sparsity of channel matrices is exploited.}
Exploiting the knowledge of the slow-varying channel components and the channel sparsity, a message-passing based algorithm is proposed in \cite{LiuJSAC2020} to estimate the cascaded channels. The work in \cite{WangTWC2020} proposed a three-phase pilot-based channel estimation framework for RIS-assisted uplink multiuser communications. A sparsity-structured tensor decomposition-based channel estimation method is proposed in \cite{ZhangCL2022}. {However, the complexity of aforementioned works grows rapidly with the square or cube of the number of RIS elements or has special requirements on the relevant matrices (e.g., the phase matrix of RIS), hindering their applications.} 

{In this letter, to fully exploit the structure of the channels in RIS-aided mmWave MIMO communications, we formulate the channel estimation as a hierarchically destructed matrix recovery problem. To solve the problem, we develop a Bayesian method, which is efficiently implemented with low complexity leveraging unitary approximate message passing (UAMP) \cite{2015GuoApproximate,YuanTSP2021} and the fast Fourier transform (FFT).  %Moreover, it exhibits high robustness to the relevant matrices. 
Simulation results show that the proposed channel estimator can achieve significant performance gain and delivers performance close to the oracle bound.}

%we propose a novel RIS channel estimation algorithm by combining UAMP and SBL \cite{YuanTSP2021,UAMPSBL,2015GuoApproximate}. Specifically, by utilizing the sparsity structure of mmWave MIMO channels in the angle domain, we derive the proposed algorithm by applying UAMP and SBL. The proposed algorithm can be implemented in a fast way via the Fast Fourier Transform (FFT). Moreover, \rev{it exhibits high robustness to the relevant matrices. Compared with state-of-the-art algorithms, the proposed channel estimator achieves performance enhancement and delivers performance closer to the oracle bound.}

\textit{Notations}-Boldface lower-case and upper-case letters denote vectors and matrices, respectively. Superscripts ${(\boldsymbol{A})}^{H}$, ${(\boldsymbol{A})}^{*}$and ${(\boldsymbol{A})}^{T}$ represent conjugate transpose, conjugate and transpose of $\boldsymbol{A}$, respectively. $[\boldsymbol{A}]_{i,j}$ represents the ($i,j$)-th element of $\boldsymbol{A}$'s, while $[\boldsymbol{A}]_{i,:}$ and $[\boldsymbol{A}]_{:,j}$ represent its $i$-th row and $j$-th column, respectively. A Gaussian distribution of $x$ with mean $\hat{x}$ and variance $\nu_{x}$ is denoted by $\mathcal{N}(x ; \hat{x}, \nu_{x}).$ Notations $\otimes$ and $\odot$ represent the Kronecker and  Khatri-Rao products, respectively. The notation $\langle\boldsymbol{a}\rangle$ denotes an average operation and $\boldsymbol{I}_{n}$ denotes a $n \times n$ identity matrix. 

\section{Channel and System Models} \label{sec:model} 
%\subsection{System Model}
We consider a RIS-aided mmWave MIMO uplink system, where a RIS with $N$ passive reflecting elements is equipped between an $M$-antenna BS and $K$ single-antenna users. We neglect the direct propagation path between the BS and users due to the high attenuation caused by unfavorable propagation environments \cite{WangSPL2020}. The channel matrices of BS-RIS and RIS-users are denoted by  $\boldsymbol{G} \in \mathbb{C}^{M \times N}$ and $\boldsymbol{H}\triangleq[\boldsymbol{h}_{1},\ldots,\boldsymbol{h}_{K}]\in \mathbb{C}^{N \times K}$, respectively. 
A narrow band geometric channel model is used to characterize the channels $\boldsymbol{G}$ and $\boldsymbol{h}_{k}$ \cite{WangSPL2020,LiuJSAC2020}.
A uniform linear array (ULA) is adopted at the BS, and {the RIS is an $N_{1}\times N_{2}$ uniform planar array (UPA) with $N=N_{1}N_{2}$ elements.} Specifically, $\boldsymbol{G}$ can be expressed as
\begin{align}
	\boldsymbol{G}=\sqrt{{\rho}(MN)^{-1}}\sum\nolimits_{p=1}^{P}\zeta_{p}\boldsymbol{a}_{\mathrm{B}}(\psi _{p})\boldsymbol{a}_{\mathrm{R}}^{H}(\vartheta _{p},\gamma _{p}), \label{eq:H_azi}
\end{align}where {$\rho$ denotes the average path loss, $P$ is the number of paths, $\zeta_{p}$ represents the complex gain associated with the $p$-th path, $\psi _{p}$ denotes the corresponding azimuth angle-of-arrival (AoA), and $ \vartheta _{p}$ and $\gamma _{p}$ respectively denote the associated azimuth and elevation angle-of-departure (AoD). {The two vectors} $\boldsymbol{a}_{\mathrm{B}}(\phi _{p})$ and $\boldsymbol{a}_{\mathrm{R}}(\vartheta _{p},\gamma _{p})$ are the array response vector of the BS and RIS, respectively, which are given by
\begin{align}
	\boldsymbol{a}_{\mathrm{B}}(\psi _{p})&=\boldsymbol{e}_{M}(\sin(\psi_{p})),\\
	\boldsymbol{a}_{\mathrm{R}}(\vartheta_{p},\gamma_{p})&=\boldsymbol{e}_{N_1}(\cos(\gamma_{p}))
	\otimes \boldsymbol{e}_{N_2}(\sin(\gamma_{p})\cos(\vartheta_{p})),\label{eq:a_RIS}
\end{align}
where $\boldsymbol{e}_{M}(x)\triangleq {M^{-\frac{1}{2}}}[1,e^{-j\frac{2\pi d}{\lambda}x},\cdots,e^{-j\frac{2\pi d}{\lambda}(M-1)x} ]^{T}$, $\lambda$ and $d$ respectively denote the wavelength and antenna spacing with $\lambda=2d$. The RIS-BS channel has a sparse representation in the angular domain, i.e.,
\begin{align}\label{eq:G}
	\boldsymbol{G}=\boldsymbol{F}_{1}\boldsymbol{\Omega}(\boldsymbol{F}_{x}\otimes\boldsymbol{F}_{y})^{H}\triangleq \boldsymbol{F}_{1}\boldsymbol{\Omega}\boldsymbol{F}_{2}^{H},
\end{align}
where $ \boldsymbol{F}_{1} \in \mathbb{C}^{M \times M}$, $\boldsymbol{F}_{x} \in \mathbb{C}^{N_{1} \times N_{1}}$ and $\boldsymbol{F}_{y} \in \mathbb{C}^{N_{2} \times N_{2}}$ (and $\boldsymbol{F}_{2} \in \mathbb{C}^{N \times N}$) are unitary Discrete Fourier Transform (DFT) matrices, $\boldsymbol{\Omega}\triangleq [\boldsymbol{\omega}_1,\cdots,\boldsymbol{\omega}_N]\in \mathbb{C}^{M \times N}$ is a sparse matrix with $P$ non-zero entries corresponding to the channel path gains $\zeta_{p}$, where $\boldsymbol{\omega}_n=[{\omega}_{1,n},\cdots,{\omega}_{M,n}]^T\in \mathbb{C}^{M \times 1}$.
%$\boldsymbol{F}_{B}=[\boldsymbol{a}(\psi _{1},L),\cdots\,\boldsymbol{a}(\psi _{{L}^{\prime}},L)]\in\mathbb{C}^{L\times {L}^{\prime}}$ is an overcomplete matrix $({L}^{\prime}\geq L)$, {in which each column is a steering vector chosen from} a pre-discretized AoA. $\boldsymbol{F}_{x}\triangleq [\boldsymbol{a}(\omega _{1},N_{2}),\cdots,\boldsymbol{a}(\omega_{{N_{2}}^{\prime}},N_{2})]\in\mathbb{C}^{N_{2}\times {N_{2}^{\prime}}}$ and $\boldsymbol{F}_{y}\triangleq [\boldsymbol{a}(\theta _{1},N_{1}),\cdots,\boldsymbol{a}(\theta _{{N_{1}}^{\prime}},N_{1})]\in\mathbb{C}^{N_{1}\times {N_{1}^{\prime}}}$ also are overcomplete matrices (${N_{2}}^{\prime}\geq N_{2}, {N_{1}}^{\prime}\geq N_{1} $). {They are defined similarity with $\boldsymbol{F}_{B}$, that is, each column is a steering vector chosen from a pre-discretized AoD.}  Since the total number of paths $P$ is small, {$\boldsymbol{D}\in\mathbb{C}^{{L}^{\prime}\times {N}^{\prime}}$ is a sparse matrix, where the  $P$ non-zero elements are equal to the channel path gains $\left\lbrace \varrho_{1},\cdots,\varrho_{P}\right\rbrace $.} Here for simplicity, we assume that the true AoA and AoD parameters lie on the discretized grids.
Similarly, the channel between the $k$-th user and RIS $\boldsymbol{h}_{k}\triangleq [{h}_{1,k},\cdots,{h}_{N,k}]^T$ can be modeled as
\begin{align}
	\boldsymbol{h}_{k}=\sqrt{{\xi _{k}}N^{-1} }\sum\nolimits_{p=1}^{{P}^{\prime}}\lambda _{p}\boldsymbol{a}_{\mathrm{R}}(\vartheta _{p},\gamma _{p}),\label{eq:g_m_azi}
\end{align}where ${P}^{\prime}$ is the number of paths between the $k$-th user and the RIS, $\xi _{k}$ and $\lambda_{p}$ denote the average path-loss and the complex gain, respectively. The $k$-th user-RIS channel $\boldsymbol{h}_{k}$ can be rewritten as 
\begin{align}
\boldsymbol{h}_{k}=\boldsymbol{F}_{2}\boldsymbol{\sigma}_{k}, 
\end{align}
where $\boldsymbol{\sigma}_{k} \triangleq[{\sigma}_{1,k},\cdots,{\sigma}_{N,k}]^T \in\mathbb{C}^{N\times 1}$ is also a sparse vector with ${P}^{\prime}$ non-zero entries. Stacking $\boldsymbol{h}_{k}$ into a matrix, the channel matrix $\boldsymbol{H}$ can be rewritten as
\begin{align}
	\boldsymbol{H}=\boldsymbol{F}_{2}\boldsymbol{\Sigma},\label{eq:H}
\end{align}
where $\boldsymbol{\Sigma} \triangleq[\boldsymbol{\sigma}_{1},\cdots,\boldsymbol{\sigma}_{K}]$.

Assuming that the RIS has $L$ available phase configurations, the received signal by the BS for $T$ consecutive time slots with the $l$-th ($l =1, \ldots, L$) RIS phase configuration {is denoted by $\boldsymbol{Y}_{l} \in \mathbb{C}^{M \times T}$, which is given as}
\begin{align}
	\boldsymbol{Y}_{l} = \boldsymbol{G} \mathrm{diag} ([\boldsymbol{\Phi}]_{l,:}) \boldsymbol{H} \boldsymbol{X}+\boldsymbol{W}_{l}, \label{eq:recvYl}
\end{align}
where $\boldsymbol{\Phi} \in \mathbb{C}^{L \times N}$ is the RIS phase {matrix,} $\boldsymbol{X}\in \mathbb{C}^{K \times T}$ denotes the transmitted orthogonal training matrix from the users, i.e., $\boldsymbol{X}\boldsymbol{X}^H=\boldsymbol{I}_{K}$, {and $\boldsymbol{W}_{l}$ represents the zero mean complex additive white Gaussian noise (AWGN) with precision $\beta$.}
{Right-multiplying $\boldsymbol{X}^H$ at the both sides of  $\eqref{eq:recvYl}$ and vectorizing the processed signal lead to}
\begin{align}
	\mathrm{vec} ( \boldsymbol{\tilde{Y}}_{l}) &= ( \boldsymbol{H}^T  \otimes \boldsymbol{G}) \mathrm{vec}\left(\mathrm{diag} ([\boldsymbol{\Phi}]_{l,:} )\right)+\mathrm{vec}( \boldsymbol{\tilde{W}}_{l}),\nonumber \\
	&=(\boldsymbol{H}^T  \odot \boldsymbol{G}) ([\boldsymbol{\Phi}]_{l,:})^{T}+\mathrm{vec}( \boldsymbol{\tilde{W}}_{l}),
	\label{eq:vecYl_nopilot}
\end{align}
where $\boldsymbol{\tilde{Y}}_{l} \triangleq \boldsymbol{Y}_{l} \boldsymbol{X}^H$ and $\boldsymbol{\tilde{W}}_{l} \triangleq \boldsymbol{W}_{l} \boldsymbol{X}^H$. 
By stacking \eqref{eq:vecYl_nopilot} into a matrix-form, we have
$\boldsymbol{\tilde{Y}}=(\boldsymbol{H}^T  \odot \boldsymbol{G}) \boldsymbol{\Phi}^{T}+\boldsymbol{\tilde{W}}$,
which can be rewritten as
\begin{equation}
	\boldsymbol{Y}=\boldsymbol{\Phi}\boldsymbol{S}+\boldsymbol{W}, \label{eq:Y_nopilot}
\end{equation}
where $\boldsymbol{Y}=\boldsymbol{\tilde{Y}}^{T}$, $\boldsymbol{W}=\boldsymbol{\tilde{W}}^{T}$ and 
\begin{equation}
\boldsymbol{S}=(\boldsymbol{H}^T  \odot \boldsymbol{G})^T.	\label{eq:S}
\end{equation}

%It is noted that $L$ configurations of the RIS are used during the training process, based which we aim to estimate the channel matrices. In order to reduce the training overhead and communication latency, a small $L$ is desirable. 

{Our aim is to estimate the channel matrices based on the observation model \eqref{eq:Y_nopilot} and the sparsity constraints on the channel matrices in \eqref{eq:G} and \eqref{eq:H}. This is a hierarchically structured signal recover problem, where $\boldsymbol{S}$ admits the structure in \eqref{eq:S} with components $\boldsymbol{G}$ and  $\boldsymbol{H}$, and $\boldsymbol{G}$ and  $\boldsymbol{H}$  admit the sparsity structures in \eqref{eq:G} and \eqref{eq:H}, respectively. In this work, we will solve the problem using the Bayesian approach and develop a low-complexity message passing based algorithm.}
%Now we can see from \eqref{eq:Y_nopilot} that the channel estimation is reformulated as the recovery of the matrix $\boldsymbol{S}$. We can also treat $\boldsymbol{S}$ as an intermediate variable, as our aim is to estimate $\boldsymbol{H}$ and $\boldsymbol{G}$. 

\section{Probabilistic Formulation and Factor Graph Representation}
{Considering that the matrices $\boldsymbol{\Omega}$ and $\boldsymbol{\Sigma}$ are sparse, we adopt the sparsity promoting two-layer Gaussian-Gamma prior for them, i.e.,  
\begin{align}\label{prior}
	p(\boldsymbol{\Omega}|\boldsymbol{\Gamma}^{g})%\prod_{m,n} %p(\omega_{m,n}\!\!\mid\! \!{\gamma}^{g}_{m,n})\!
	&=\! \prod_{m} \prod_{n}\mathcal{N}(\omega_{m,n}; 0,({\gamma}^{g}_{m,n})^{-1}),  \\
	p(\boldsymbol{\Gamma}^{g})&=\prod_{m} \prod_{n}Ga({\gamma}^{g}_{m,n}; \epsilon^{g}_{m,n}, \eta^{g}_{m,n}),\\
	p(\boldsymbol{\Sigma}\mid \boldsymbol{\Gamma}^{h})&= \prod_{n}\prod_{k}  \mathcal{N}(\sigma_{n,k}; 0,({\gamma}^{h}_{n,k})^{-1}), \\
	p(\boldsymbol{\Gamma}^{h})&= \prod_{n}\prod_{k}  Ga({\gamma}^{h}_{n,k}; \epsilon^{h}_{n,k}, \eta^{h}_{n,k}),
\end{align}
where $\boldsymbol{\Gamma}^{g}=[\boldsymbol{\gamma}^{g}_{1},\cdots,\boldsymbol{\gamma}^{g}_{N}] $ ($\boldsymbol{\gamma}^{g}_{n} \triangleq[{\gamma}^{g}_{1,n},\cdots,{\gamma}^{g}_{M,n}]^T$) and $\boldsymbol{\Gamma}^{h}= [\boldsymbol{\gamma}^{h}_{1},\cdots,\boldsymbol{\gamma}^{h}_{K}] $ ($\boldsymbol{\gamma}^{h}_{k} \triangleq[{\gamma}^{h}_{1,k},\cdots,{\gamma}^{g}_{N,k}]^T$).} 
{To handle the dense matrix $\boldsymbol{\Phi}$ in \eqref{eq:Y_nopilot} in designing our message passing algorithm, we employ UAMP, where with the singular value decomposition (SVD) for matrix $\boldsymbol{\Phi}$, i.e., $\boldsymbol{\Phi}=\boldsymbol{U \Lambda V}$, a unitary transformation to \eqref{eq:Y_nopilot} is performed \cite{2015GuoApproximate,YuanTSP2021, UAMPSBL}, i.e.,}
\begin{align}\label{svd}
\boldsymbol{R}=\boldsymbol{\Psi}\boldsymbol{S}+\boldsymbol{\bar{W}},
\end{align}
where $\boldsymbol{R}=\boldsymbol{U}^{H} \boldsymbol{Y}$, $\boldsymbol{\Psi}=\boldsymbol{U}^{H} \boldsymbol{\Phi}=\boldsymbol{\Lambda V}$, and $\boldsymbol{\bar{W}}=\boldsymbol{U}^{H} \boldsymbol{W}$ remains a zero-mean Gaussian noise with the same precision $\beta$. Let $J\!\!=\!\!KM$, and note that $\boldsymbol{R}\!\!=\!\![\boldsymbol{r}_{1}, \ldots, \boldsymbol{r}_{J}]$ ($\boldsymbol{r}_{j}\triangleq[{r}_{1,j},\cdots,{r}_{L,j}]^T$), $\boldsymbol{S}\!\!=\!\![\boldsymbol{s}_{1},\ldots,\boldsymbol{s}_{J}]$ ($\boldsymbol{s}_{j}\triangleq[{s}_{1,j},\cdots,{s}_{N,j}]^T$), $\boldsymbol{\bar{W}}=[\boldsymbol{w}_{1},\ldots,\boldsymbol{w}_{J}]$ and define the auxiliary variable $\boldsymbol{Z}\triangleq[\boldsymbol{z}_{1},\ldots,\boldsymbol{z}_{J}]$ ($\boldsymbol{z}_{j}\triangleq[{z}_{1,j},\cdots,{z}_{L,j}]^T$) with $\boldsymbol{z}_{j}=\boldsymbol{\Psi}\boldsymbol{s}_{j}$, i.e., $\boldsymbol{r}_{j}=\boldsymbol{z}_{j}+\boldsymbol{w}_{j}$.

\begin{figure*}[h]
	\centering
	\includegraphics[width=1.1\columnwidth]{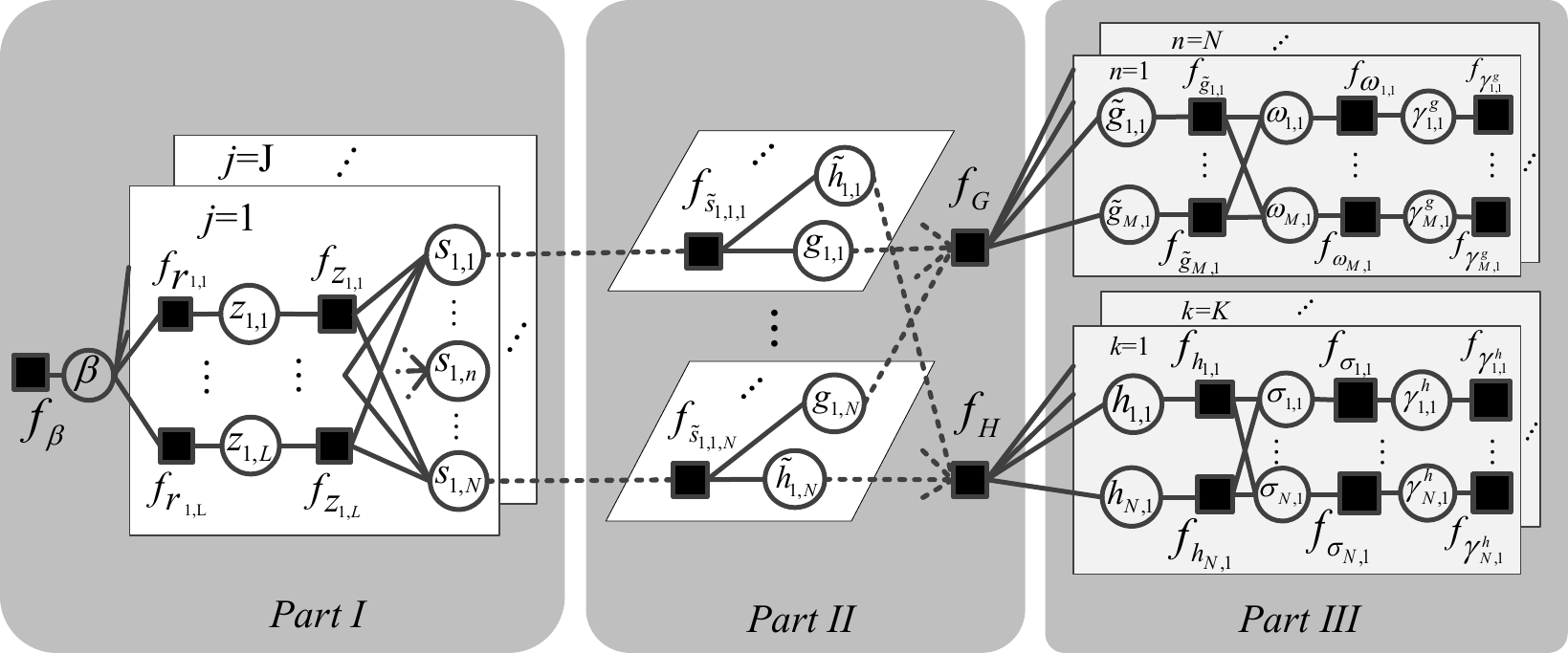}
	\caption{{ Factor graph representation of \eqref{eq:Factor}.}} \label{fig:FactorGraph}
\end{figure*}

The relationship $\boldsymbol{{S}}^T=\boldsymbol{{H}}^T \odot \boldsymbol{G}$ in \eqref{eq:S} indicates $\boldsymbol{\tilde{s}}_{n}=\boldsymbol{\tilde{h}}_{n} \otimes \boldsymbol{g}_{n}$ with $\boldsymbol{{S}}^T\triangleq [\boldsymbol{\tilde{s}}_{1},\ldots,\boldsymbol{\tilde{s}}_{N}]$, $\boldsymbol{{H}}^T \triangleq [\boldsymbol{\tilde{h}}_{1},\ldots,\boldsymbol{\tilde{h}}_{N}]$ and $\boldsymbol{G}\triangleq[\boldsymbol{g}_{1},\ldots,\boldsymbol{g}_{N}]$, or in scalar form, i.e, ${\tilde{s}}_{m,k,n}=\tilde{h}_{k,n}{g}_{m,n}$ with $\boldsymbol{\tilde{s}}_{n}\triangleq[{\tilde{s}}_{1,1,n},\ldots,{\tilde{s}}_{m,k,n},\ldots,{\tilde{s}}_{M,K,n}]^T$, $\boldsymbol{\tilde{h}}_{n}\triangleq[\tilde{h}_{1,n},\ldots,\tilde{h}_{K,n}]^T$ and $\boldsymbol{g}_{n}\triangleq[{g}_{1,n},\ldots,{g}_{M,n}]^T$. We next define an auxiliary variable $\boldsymbol{\tilde{G}} \triangleq \boldsymbol{F}_{1}\boldsymbol{\Omega}$ with $\boldsymbol{\tilde{G}}\triangleq[\boldsymbol{\tilde{g}}_{1},\cdots,\boldsymbol{\tilde{g}}_{N}]$ and $\boldsymbol{\tilde{g}}_{n}\triangleq[{\tilde{g}}_{1,n},\cdots,{\tilde{g}}_{M,n}]^T$.
Then, we have the following joint distribution
\begin{align}
	p&(\boldsymbol{H}, \boldsymbol{G}, \boldsymbol{\tilde{G}}, \boldsymbol{S}, \boldsymbol{Z},\boldsymbol{\Omega},\boldsymbol{\Sigma}, \boldsymbol{\Gamma}^{g}, \boldsymbol{\Gamma}^{h},  \beta |\boldsymbol{R}) \nonumber \\ 
%	\propto 
%	&  ~p(\beta) p(\boldsymbol{R}|\boldsymbol{Z},\beta)
%	p(\boldsymbol{Z}|\boldsymbol{S})
%	p(\boldsymbol{{S}}|\boldsymbol{H},\boldsymbol{G}) 
%	p(\boldsymbol{{G}}|\boldsymbol{\tilde{G}})
%	p(\boldsymbol{\tilde{G}}|\boldsymbol{\Omega})
%	\nonumber \\ &
%	 p(\boldsymbol{\Omega}|\boldsymbol{\Gamma}^g)	 
%	 p(\boldsymbol{\Gamma}^g) 
%	 p(\boldsymbol{H}|\boldsymbol{{H}}^{T})
%     p(\boldsymbol{{H}}^{T}|\boldsymbol{\Sigma})	 
%    	 p(\boldsymbol{\Sigma}|\boldsymbol{\Gamma}^h)
%    	  	{p(\boldsymbol{\Gamma}^h)}
%	 \nonumber \\
	= &
    f_{\beta}(\beta)
	{\prod}_{l}{\prod}_{j}  
	\left[  f_{{r}_{l,j}}({r}_{l,j},{z}_{l,j},\beta)	
	f_{{z}_{l,j}}({z}_{l,j},\boldsymbol{s}_{j})\right]      
	\nonumber \\	&  ~
	\times
	{\prod}_{m}{\prod}_{n}
\Big[  {\prod}_{k} \left(  f_{{\tilde{s}}_{m,k,n}}({\tilde{s}}_{m,k,n},\tilde{h}_{k,n},{g}_{m,n}) \right) 
	\nonumber \\&  ~
	f_{\boldsymbol{G}}({{g}}_{m,n}, [\boldsymbol{\tilde{G}}]_{m,:})
   	f_{{\tilde{g}}_{m,n}}({\tilde{g}}_{m,n},{\omega}_{m,n} ) 
    f_{{\omega}_{m,n}}({\omega}_{m,n},{\gamma}^g_{m,n} ) 
\nonumber \\&  ~
	f_{{\gamma}^g_{m,n}}({\gamma}^g_{m,n} ) \Big]
	\times  
	{\prod}_{n}  {\prod}_{k}   
	[ 	f_{\boldsymbol{H}}(\tilde{h}_{k,n}, {{h}}_{n,k})	
	    f_{{{h}}_{n,k}}({{h}}_{n,k},{\sigma}_{n,k})	
	\nonumber \\& ~
	f_{{\sigma}_{n,k}}({\sigma}_{n,k},{\gamma}^h_{n,k})
	f_{{\gamma}^h_{n,k}}({\gamma}^h_{n,k} )],\label{eq:Factor}
\end{align}
where the involved distributions are listed in Table \ref{tab:factor}. 
%To facilitate the factor graph representation of the factorization in \eqref{eq:Factor}, local functions (factors) are defined, and the correspondence between the distributions and local functions are also shown in Table \ref{tab:factor}. 
%\rev{We assume non-informative priors $p(\rho_h)$ and $p(\rho_g)$  for the precisions $\rho_h$ and $\rho_g$, i.e., they are uniform distributions over $0$ to $+\infty$. So the corresponding local functions in \eqref{eq:Factor} are omitted.} %It is noted that \rev{$p(\boldsymbol{H}|\rho_h)$ and  $p(\boldsymbol{G}|\rho_g)$} represent the priors for the channel matrices $\boldsymbol{H}$ and  $\boldsymbol{G}$, respectively. When no priors are available for the channel matrices, the priors can be set to be a non-informative one, e.g., \rev{$\rho_h^{-1}=\rho_g^{-1}= +\infty$} in Table \ref{tab:factor}. 
The factor graph representation of \eqref{eq:Factor} is depicted in Fig. \ref{fig:FactorGraph}. Based on Fig. \ref{fig:FactorGraph}, we will {develop a low-complexity message passing based algorithm to obtain the approximate marginals about the entries of  \eqref{eq:G} and \eqref{eq:H}, thereby their estimates.}
\begin{table}[htb]\footnotesize
	\color{black} 
	\centering
	\renewcommand\arraystretch{1.2}
	\caption{Factors and distributions in (\ref{eq:Factor}).}\label{tab:factor}
	\begin{tabular}{>{ }p{25pt}>{ }p{40pt} >{ \arraybackslash }p{130pt}}
		\hline
		Factor & Distribution & Function  \\
		\hline
		$ f_{\beta}$  & $p(\beta)$ & $\propto \beta^{-1}$ \\		
		$f_{{r}_{l,j}}$ & $p(\boldsymbol{R}|\boldsymbol{Z},\beta)$ & ${\prod}_{l}  {\prod}_{j} \mathcal{N}({r}_{l,j};{z}_{l,j}, \beta^{-1})$ \\		
		$f_{{z}_{l,j}}$ & $p(\boldsymbol{Z}|\boldsymbol{S})$ & ${\prod}_{l}  {\prod}_{j}\delta({z}_{l,j}-[\boldsymbol{\Psi}]_{l,:}\boldsymbol{s}_{j})$ \\
		$f_{{\tilde{s}}_{m,k,n}}$ &	$p(\boldsymbol{{S}}|\boldsymbol{H},\boldsymbol{G})$ & ${\prod}_{m} {\prod}_{k} {\prod}_{n}\delta({\tilde{s}}_{m,k,n}\!-\!\tilde{h}_{k,n} \otimes {g}_{m,n})$ \\			
		$f_{\boldsymbol{G}}$ &	$p(\boldsymbol{{G}}|\boldsymbol{\tilde{G}})$ & ${\prod}_{m} {\prod}_{n}\delta({{g}}_{m,n}-{[\boldsymbol{\tilde{G}}]}_{m,:}{[\boldsymbol{F}^{H}_2]}_{:,n})$ \\	
		$f_{{\tilde{g}}_{m,n}}$ &	$p(\boldsymbol{\tilde{G}}|\boldsymbol{\Omega})$ & ${\prod}_{m} {\prod}_{n}\delta({\tilde{g}}_{m,n}-{[\boldsymbol{F}_1]}_{m,:}{\boldsymbol{\omega}}_{n})$ \\	
		$f_{{\omega}_{m,n}}$ & $p(\boldsymbol{\Omega}|\boldsymbol{\Gamma}^g)$ & ${\prod}_{m}  {\prod}_{n} \mathcal{N}({\omega}_{m,n};0, (\gamma^{g}_{m,n})^{-1})$ \\		
		$f_{{\gamma}^{g}_{m,n}}$ & $p(\boldsymbol{\Gamma}^g)$ & ${\prod}_{m}  {\prod}_{n} {Ga}(\gamma^{g}_{m,n};\epsilon^{g}_{m,n},\eta^{g}_{m,n} )$ \\	
		$f_{\boldsymbol{H}}$ &	$p(\boldsymbol{H}|\boldsymbol{{H}}^T)$ & ${\prod}_{n} {\prod}_{k}\delta(\tilde{h}_{k,n}-{{h}}_{n,k})$ \\			
		$f_{{{h}}_{n,k}}$ &	$p(\boldsymbol{{H}}^T|\boldsymbol{\Sigma})$ & ${\prod}_{n} {\prod}_{k}\delta({{h}}_{n,k}-{[\boldsymbol{F}_2]}_{n,:}{\boldsymbol{\sigma}}_{k})$ \\	
		$f_{{\sigma}_{n,k}}$ & $p(\boldsymbol{\Sigma}|\boldsymbol{\Gamma}^h)$ & ${\prod}_{n}  {\prod}_{k} \mathcal{N}({\sigma}_{n,k};0, (\gamma^{h}_{n,k})^{-1})$ \\		
        $f_{{\gamma}^{h}_{n,k}}$ & $p(\boldsymbol{\Gamma}^h)$ & ${\prod}_{n}  {\prod}_{k} {Ga}(\gamma^{h}_{n,k};\epsilon^{h}_{n,k},\eta^{h}_{n,k} )$ \\	
		\hline
	\end{tabular}
\end{table}

\section{Message Passing Algorithm Design}

%In this section, we apply the advanced message passing techniques UAMP and SBL in Fig. 1 to recover the sparse mmWave channels with relative low complexity. For simplicity of description, 
{As shown in Fig. 1, we divide the factor graph into three parts, and elaborate the message computations in each part.}  %the forward and backward derivation is detailed below.

\subsubsection{Message Computations in Part $\uppercase\expandafter{\romannumeral1}$}
{Due to the dense connections in this part and the non-i.i.d. Gaussian entries of matrix $\boldsymbol{\Phi}$, UAMP is used to handle the message passing in this part. Here we borrow the algorithm developed in \cite{GuoTWC2023}.
%The intensive connections exist in Part $\uppercase\expandafter{\romannumeral1}$ of Fig. \ref{fig:FactorGraph}, where we can turn to UAMP for help. 
Due to the space limitation, we do not provide the details, which can be found in \cite{GuoTWC2023}.} However, it is necessary to illustrate the incoming message $m_{f_{{\tilde{s}}_{m,k,n}}\rightarrow {{s}}_{j,n}}( {{s}}_{j,n})$ and outgoing message $m_{{s}_{j,n} \rightarrow f_{\tilde{s}_{m,k,n}}}({s}_{j,n})$. 

Part $\uppercase\expandafter{\romannumeral2}$ feeds the incoming message to UAMP as the priori of ${s}_{j,n}$, i.e., $ m_{f_{{\tilde{s}}_{m,k,n}}\rightarrow {{s}}_{j,n}}( {{s}}_{j,n})= \mathcal{N}(s_{j,n};\larrow{s}_{j,n},\larrow{\nu}_{{s}_{j,n}})$, where $\larrow{\nu}_{{s}_{j,n}}$ and $\larrow{s}_{j,n}$ can be respectively updated as 
 \begin{align}
	\larrow{s}_{j,n}&=[\larrow{\boldsymbol{g}}_{n} \cdot \larrow{\tilde{\boldsymbol{h}}}_{n}]_j,\label{s_jn}\\
	\larrow{\nu}_{{s}_{j,n}}\!\!=\![|\larrow{\tilde{\boldsymbol{h}}}_{n}|^2 \cdot\larrow{\boldsymbol{\nu}}_{\boldsymbol{g}_{n}}\!\!&+\!|\larrow{\boldsymbol{g}}_{n}|^2\cdot \larrow{\boldsymbol{\nu}}_{\tilde{\boldsymbol{h}}_{n}}\!+\!\larrow{\boldsymbol{\nu}}_{\tilde{\boldsymbol{h}}_{n}} \cdot \larrow{\boldsymbol{\nu}}_{\boldsymbol{g}_{n}}]_j,\label{le_s_jn}
\end{align}	
where $ \larrow{\boldsymbol{\nu}}_{\boldsymbol{g}_{n}} \triangleq [\larrow{\nu}_{g_{1,1,n}},\cdots,\larrow{\nu}_{g_{M,K,n}}]^T$, $\larrow{\boldsymbol{g}}_{n}\triangleq[ \larrow{g}_{1,1,n},\cdots, \larrow{g}_{M,K,n}]^T$, $\larrow{\boldsymbol{\nu}}_{\boldsymbol{g}_{n}}$, $\larrow{\boldsymbol{g}}_{n}$, $\larrow{\boldsymbol{\nu}}_{\tilde{\boldsymbol{h}}_{n}}$ and $\larrow{\tilde{\boldsymbol{h}}}_{n}$ can be computed in the previous iteration.
%	by \eqref{le_s_jn} and \eqref{s_jn}.
%Leveraging the UAMP algorithm \cite{2015GuoApproximate,2020YuanApproximate}, we can derive the involved message passing process of Part $\uppercase\expandafter{\romannumeral1}$. However, due to space limitations, we have to only show the outgoing message in this part's forward and backward message passing. We first compute the forward message $\{m_{{s}_{j,n} \rightarrow f_{\tilde{s}_{m,k,n}}}({s}_{j,n})\}$, which are input to the $\emph{Part \uppercase\expandafter{\romannumeral2}}$, i.e.,
Furthermore, UAMP output the message $m_{{s}_{j,n} \rightarrow f_{\tilde{s}_{m,k,n}}}({s}_{j,n})$ as the input of Part II, i.e.,
\begin{align}\label{P11}
m_{{s}_{j,n} \rightarrow f_{\tilde{s}_{m,k,n}}}({s}_{j,n})=\mathcal{N}({s}_{j,n};{q}_{j,n},\big \langle \sum_{n=1}^{N}{\nu}_{{q}_{j,n}}\big\rangle),
\end{align}
where ${q}_{j,n}$ and ${\nu}_{{q}_{j,n}}$ are respectively the $n$-th elements of $\boldsymbol{q}_{j}$ and $\boldsymbol{\nu}_{\boldsymbol{q}_{j}}$ given in the Line 4 of the Algorithm 1 in \cite{GuoTWC2023}. 

\subsubsection{Message Computations in Part $\uppercase\expandafter{\romannumeral2}$}
According to the deterministic relation $\boldsymbol{\tilde{G}} = \boldsymbol{F}_{1}\boldsymbol{\Omega}$, the mean ${g^{\prime}}_{m,n}$ and variance $\nu_{{g^{\prime}}_{m,n}}$ of the Gaussian message $m_{f_{\boldsymbol{G}}\rightarrow {g}_{m,n} }({g}_{m,n})$ can be computed as
\begin{align}\label{tilde_g_mn}
	\nu_{g^{\prime}_{m,n}}=\langle \boldsymbol{\nu}_{p^{g}_{n}}\rangle,~~~~
	g^{\prime}_{m,n}=[\boldsymbol{P}^{g}]_{m,:}[\boldsymbol{F}^{H}_{2}]_{:,n},
\end{align}
where $\boldsymbol{P}^{g}$ is the corresponding matrix stacked by $\{p^{g}_{m,n}\}$, $\boldsymbol{\nu}_{p^{g}_{n}}$ is the vector stacked by $\{\nu_{p^{g}_{m,n}}\}$, $\nu_{p^{g}_{m,n}}$ and $p^{g}_{m,n}$ are respectively the variance and mean of $\tilde{g}_{m,n}$ which will be computed in \eqref{nu_p_g_mn}.
So the belief of ${g}_{m,n}$ can be expressed as
\begin{align}\label{b_g_mn}
	\mathfrak{b}({g}_{m,n})&\propto m_{g_{m,n} \rightarrow f_{{\boldsymbol{G} }}}\left(g_{m,n}\right) m_{ f_{{\boldsymbol{G} }} \rightarrow g_{m,n}  }\left(g_{m,n}\right) \nonumber \\ 
	&\propto \mathcal{N} (g_{m,n};\hat{g}_{m,n},\nu_{g_{m,n}}),
\end{align}
where the message $m_{g_{m,n} \rightarrow f_{{\boldsymbol{G} }}}(g_{m,n})\!=\!\mathcal{N}(g_{m,n}; \rarrow{g}_{m,n}, \rarrow{\nu}_{g_{m,n}})$ can be updated in \eqref{right_v_g_nm} and \eqref{right_g_nm}, and
\begin{align}\label{hat_g_mn}
	\nu_{{g}_{m,n}}	\!\!=\! \rarrow{\nu}^{-1}_{g_{m,n}}\!\!+\!{\nu}^{-1}_{{g^{\prime}}_{m,n}},~~~~
	\hat{g}_{m,n}\!\!=\!\rarrow{g}_{m,n}\rarrow{\nu}^{-1}_{g_{m,n}}\!\!+\!{g^{\prime}}_{m,n}{\nu}^{-1}_{{g^{\prime}}_{m,n}}.
\end{align}
Furthermore, based on the belief propagation, the backward message $m_{{g}_{m,n}\rightarrow f_{{\tilde{s}}_{m,k,n}}}({g}_{m,n}) $ can be computed as
\begin{align}
	m_{{g}_{m,n}\rightarrow f_{{\tilde{s}}_{m,k,n}}}({g}_{m,n})=\frac{\mathfrak{b}({g}_{m,n}) }{m_{f_{{\tilde{s}}_{m,k,n}} \rightarrow {g}_{m,n} }({g}_{m,n})}.
\end{align}
with variance $\larrow{\nu}_{g_{m,k,n}}$ and mean $\larrow{g}_{m,k,n}$, which can be computed as 
\begin{align}
{\larrow{\nu}_{g_{m,k,n}}}&=[\boldsymbol{\nu}_{\boldsymbol{g}_{n}}]_{{mod(mk)}_K}^{-1}-\rarrow{\nu}^{-1}_{g_{m,k,n}}, \\
{ \larrow{g}_{m,k,n} }&=\larrow{\nu}_{g_{m,k,n}}(\hat{g}_{m,n}\nu^{-1}_{{g}_{m,n}}-\rarrow{g}_{m,k,n}\rarrow{\nu}^{-1}_{g_{m,k,n}}).
\end{align}
where $mod(i)_{K}$ denotes the $i$-modulo-$K$, $\boldsymbol{\nu}_{\boldsymbol{g}_{n}}\triangleq[\nu_{{g}_{1,n}},\cdots,\nu_{{g}_{M,n}}]^T$, $\rarrow{\nu}_{g_{m,k,n}}$ and $\rarrow{g}_{m,k,n}$ will be respectively computed in \eqref{right_v_g_nmk} and \eqref{right_g_nmk}.
%Then, we have the backward message $ m_{f_{{\tilde{s}}_{m,k,n}}\rightarrow {{s}}_{j,n}}( {{s}}_{j,n})= \mathcal{N}(s_{j,n};\larrow{s}_{j,n},\larrow{\nu}_{{s}_{j,n}})$ with
% \begin{align}\label{s_jn}
% 	\larrow{s}_{j,n}=[\larrow{\boldsymbol{g}}_{n} \cdot \larrow{\tilde{\boldsymbol{h}}}_{n}]_j,
% \end{align}
% \begin{align}\label{le_s_jn}
%	\larrow{\nu}_{{s}_{j,n}}\!\!=\![|\larrow{\tilde{\boldsymbol{h}}}_{n}|^2 \cdot\larrow{\boldsymbol{\nu}}_{\boldsymbol{g}_{n}}\!\!+\!|\larrow{\boldsymbol{g}}_{n}|^2\cdot \larrow{\boldsymbol{\nu}}_{\tilde{\boldsymbol{h}}_{n}}\!+\!\larrow{\boldsymbol{\nu}}_{\tilde{\boldsymbol{h}}_{n}} \cdot \larrow{\boldsymbol{\nu}}_{\boldsymbol{g}_{n}}]_j,
%\end{align}
%where $ \larrow{\boldsymbol{\nu}}_{\boldsymbol{g}_{n}} \triangleq [\larrow{\nu}_{g_{1,1,n}},\cdots,\larrow{\nu}_{g_{M,K,n}}]^T$, $\larrow{\boldsymbol{g}}_{n}\triangleq[ \larrow{g}_{1,1,n},\cdots, \larrow{g}_{M,K,n}]^T$, the calculations of $\larrow{\boldsymbol{\nu}}_{\tilde{\boldsymbol{h}}_{n}}$ and $\larrow{\tilde{\boldsymbol{h}}}_{n}$ can be computed in the previous iteration.

Next, with the definition $[\boldsymbol{\tilde{q}}_{1},\cdots,\boldsymbol{\tilde{q}}_{N}] \triangleq [\boldsymbol{q}_{1},\cdots,\boldsymbol{q}_{J}]^T$ and $[\boldsymbol{\nu}_{\boldsymbol{\tilde{q}}_{1}},\cdots,\boldsymbol{\nu}_{\boldsymbol{\tilde{q}}_{N}}] \triangleq [\boldsymbol{\nu}_{{\boldsymbol{q}}_{1}},\cdots,\boldsymbol{\nu}_{{\boldsymbol{q}}_{J}}]^T$, mean field rules \cite{2015GuoApproximate,YuanTSP2021} indicate that the forward message $m_{f_{\tilde{s}_{m,k,n}}\rightarrow  g_{m,n}}(g_{m,n}) \propto\mathcal{N}(g_{m,n};\rarrow{g}_{m,k,n}, \rarrow{\nu}_{g_{m,k,n}})$ with
\begin{align}
	\rarrow{\nu}_{g_{m,k,n}} &=\langle \boldsymbol{\nu}_{\boldsymbol{\tilde{q}}_{k,n}} \rangle({|\hat{\tilde{h}}_{k,n}|^{2}+\nu_{\tilde{h}_{k,n}}})^{-1},\label{right_v_g_nmk} \\
	\rarrow{g}_{m,k,n} &={\tilde{q}_{m,k,n} \hat{\tilde{h}}_{k,n}^{*}}({|\hat{\tilde{h}}_{k,n}|^{2}+\nu_{\tilde{h}_{k,n}}})^{-1},\label{right_g_nmk}
\end{align}
where $ \hat{\tilde{h}}_{k,n}$ and $\nu_{\tilde{h}_{k,n}}$ are the approximated a posteriori mean and variance of ${\tilde{h}}_{k,n}$, which are computed in the previous iteration, $\boldsymbol{\nu}_{\boldsymbol{\tilde{q}}_{k,n}}$ is the $k$-th block of  $\boldsymbol{\nu}_{\boldsymbol{\tilde{q}}_{n}} \triangleq [\boldsymbol{\nu}_{\boldsymbol{\tilde{q}}_{1,n}}^T,\cdots,\boldsymbol{\nu}_{\boldsymbol{\tilde{q}}_{K,n}}^T] ^T$, ${\boldsymbol{\tilde{q}}_{n}} \triangleq [{\boldsymbol{\tilde{q}}_{1,n}}^T,\cdots,{\boldsymbol{\tilde{q}}_{K,n}}^T] ^T$, and $\tilde{q}_{m,k,n}$ is the $m$-th element of ${\boldsymbol{\tilde{q}}_{k,n}} $, i.e., ${\boldsymbol{\tilde{q}}_{k,n}}\triangleq [\tilde{q}_{1,k,n},\cdots,\tilde{q}_{M,k,n}]^T$. 
With belief propagation rules, we have $m_{g_{m,n} \rightarrow f_{{\boldsymbol{G} }}}(g_{m,n})\!=\!\mathcal{N}(g_{m,n}; \rarrow{g}_{m,n}, \rarrow{\nu}_{g_{m,n}})$ with
\begin{align}
	\rarrow{\nu}_{g_{m,n}} &=  ( \sum_{k=1}^{K} \rarrow{\nu}^{-1}_{g_{m,k,n}} ) ^{-1},\label{right_v_g_nm}  \\
	\rarrow{g}_{m,n} &=\rarrow{\nu}_{g_{m,n}}  \sum_{k=1}^{K}\!{\rarrow{g}_{m,k,n}}{\rarrow{\nu}^{-1}_{g_{m,k,n}}}.\label{right_g_nm}
\end{align}
Then, we have the message 
$m_{f_{\boldsymbol{G}} \rightarrow{\tilde{g}}_{m,n}}(\tilde{g}_{m,n})=\mathcal{N}(\tilde{g}_{m,n}; \rarrow{\tilde{g}}_{m,n}, \rarrow{\nu}_{\tilde{g}_{m,n}})$ 
with
\begin{align}\label{right_v_z_nm}
	\rarrow{\nu}_{\tilde{g}_{m,n}}=\frac{1}{N}\sum\nolimits_{n=1}^{N}\rarrow{\nu}_{g_{m,n}}^{-1},~~~~~\rarrow{\tilde{g}}_{m,n} =[ \rarrow{\boldsymbol{G}}] _{m,:}[{\boldsymbol{F}_2}]_{:,n},
\end{align}
where $ \rarrow{\boldsymbol{G}}$ is the corresponding matrix stacked by $\rarrow{\nu}_{g_{m,n}}$. 
%Then, the Gaussian form of the message $m_{f_{\boldsymbol{G}} \rightarrow{\tilde{g}}_{m,n}}(\tilde{g}_{m,n})$ suggests the following pseudo observation model
%\begin{align}\label{pseudoG}
%	\rarrow{\tilde{g}}_{m,n}& ={\tilde{g}}_{m,n} +\lambda_{m,n}\nonumber \\ 
%	&=[ {\boldsymbol{F}}_1] _{m,:}{\boldsymbol{\omega}}_{n}+\lambda_{m,n}.
%\end{align}where $\lambda_{m,n}$ is a Gaussian noise with mean $0$ and variance $ \rarrow{\nu}_{\tilde{g}_{m,n}}$. 
%The corresponding vector-form is $\rarrow{\tilde{\boldsymbol{z}}_{n}^{g}}=\boldsymbol{F}_1 \boldsymbol{\omega}_{n} +\boldsymbol{\lambda}_{n}$, where $\boldsymbol{\lambda}_{n}\triangleq[\lambda_{1,n} ,\cdots, \lambda_{M,n}]^T$. 
%This fits the forward recursion of the UAMP-SBL algorithm with the known noise variance, which will be detailed in the next part.  

Because of the symmetry between $\{\tilde{h}_{k,n}\}$ and $\{g_{m,n}\}$, the forward and backward messages related to $\{\tilde{h}_{k,n}\}$ in this part can be reached following the same steps from \eqref{tilde_g_mn} and \eqref{right_v_z_nm} by replacing the corresponding parameters, except for the following means and variables. % \footnote{The corresponding calculations related to $\{g_{m,n}\}$ are in \eqref{tilde_g_mn} and \eqref{right_v_z_nm}.}
%We will directly give these messages about $\{\tilde{h}_{k,n}\}$, i.e.,\\
%\noindent
%\emph{Backward}:
%$m_{f_{\boldsymbol{H}}\rightarrow {\tilde{h}}_{k,n} }({\tilde{h}}_{k,n})  \propto \mathcal{N}(\tilde{h}_{k,n}; {{h}^{\prime}}_{k,n},\nu_{{{h}^{\prime}}_{k,n}})$ with 
\begin{align}
	\nu_{{h}^{^{\prime}}_{k,n}}=\nu_{p^h_{n,k}},~~~~
	{{h}^{\prime}}_{k,n}=p^{h}_{n,k},\label{tilde_hnk} \\
	\rarrow{\nu}_{{h}_{n,k}}=\rarrow{\nu}_{\tilde{h}_{k,n}},~~~~~
	\rarrow{{h}}_{n,k} = \rarrow{\tilde{h}} _{k,n}.\label{z_h_nk}
\end{align}
%Note that, if a backward computation requires forward messages, the relevant messages in the previous iteration is used by default.
%This is the end of the message update in the \emph{Part} $\uppercase\expandafter{\romannumeral2}$. 
Then, we treat $m_{f_{\boldsymbol{G}} \rightarrow{\tilde{g}}_{m,n}}(\tilde{g}_{m,n})$ and $m_{f_{\boldsymbol{H}} \rightarrow{\tilde{h}}_{n,k}}(\tilde{h}_{n,k})$ as the inputs of the \emph{Part} $\uppercase\expandafter{\romannumeral3}$. 

\subsubsection{Message Computations in Part $\uppercase\expandafter{\romannumeral3}$}
We first elaborate the backward message passing. 
The message $m_{f_{\gamma^g_{m,n}} \rightarrow \gamma^g_{m,n}}(\gamma^g_{m,n}) $ is predefined Gamma distribution with shape parameter $\epsilon^{g}_{m,n}$ and rate parameter $\eta^{g}_{m,n}$, i.e.,
	\begin{align}\label{lef_gamma_mn}
		m_{ f_{\gamma^g_{m,n}}\!\! \rightarrow \gamma^g_{m,n} }(\gamma^g_{m,n}) \!\propto\!  (\gamma^g_{m,n})^{\epsilon^g_{m,n}\!-\!1}\! \exp\left\{ -\eta^g_{m,n} \gamma^g_{m,n}\right\}.
	\end{align}
So $\mathfrak{b}({\gamma}^{g}_{m,n})  \propto m_{f_{{\gamma}^{g}_{m,n}} \rightarrow {\gamma}^{g}_{m,n}}({\gamma}^{g}_{m,n})
m_{f_{{\omega}_{m,n}} \rightarrow {\gamma}^{g}_{m,n}}({\gamma}^{g}_{m,n})$. Then $\hat{\gamma}^{g}_{m,n}$ can be computed as
	\begin{align}\label{gamma_hat}
		\hat{\gamma}^{g}_{m,n}\!\!=\!\! \int \! \!\gamma^{g}_{m,n} \mathfrak{b}\left(\gamma^{g}_{m,n}\right) d\gamma^{g}_{m,n}\!\!=\!\!\frac{2\epsilon^{g}_{n}\!\!+\!1}{|\hat{\omega}_{m,n}|^2\!\!+\!\nu_{\omega_{m,n}}\!\!+\!2{\eta}^{g}_{m,n}},
	\end{align}
where $\hat{{\omega}}_{m,n}$ and ${\nu}_{{\omega }_{m,n}}$ are given in \eqref{v_gamma_nm}, and the shape parameter is tuned automatically using the rule in \cite{UAMPSBL}
\begin{align}\label{epsilon_g}
	\epsilon^{g}_{n}=0.5\sqrt{ {\log({M^{-1}}\!\sum\nolimits_{m}\hat{\gamma}^g_{m,n} )\!-\!{M^{-1}}\!\log\sum\nolimits_{m}\hat{\gamma}^g_{m,n} }}. % ^{\frac{1}{2}}.
\end{align}Then, it holds that
\begin{align}\label{omega_mn}
	m_{ f_{\omega_{m,n}} \rightarrow \omega_{m,n} }(\omega_{m,n}) \propto \mathcal{N}(\omega_{m,n};0,(\hat{\gamma}^g_{m,n})^{-1}). 
\end{align}
We then sent $m_{ f_{\omega_{m,n}} \rightarrow \omega_{m,n} }(\omega_{m,n})$ to UAMP as the prior of $\omega_{m,n}$, and UAMP outputs the Gaussian messages $ m_{ \tilde{g}_{m,n} \rightarrow f_{\boldsymbol{{G}}} } (\tilde{g}_{m,n})  = m_{f_{\tilde{g}_{m,n}} \rightarrow {\tilde{g}_{m,n}}}  \!(\tilde{g}_{m,n})$, i.e.,
\begin{align}
	{ m_{ \tilde{g}_{m,n} \rightarrow f_{\boldsymbol{{G}}} } \!(\tilde{g}_{m,n}) \!\! =  \mathcal{N}(\tilde{g}_{m,n};p^{g}_{m,n},\nu_{p^{g}_{m,n}} ) },
\end{align}
where $\nu_{p^{g}_{m,n}}$ and $p^{g}_{m,n}$ are updated as
\begin{align}
	\nu_{p^{g}_{m,n}}&=\big\langle{\sum_{m=1}^{M}{\nu}_{\omega_{m,n}}}\big\rangle,~~ 
	{p^{g}_{m,n}}\!\!=\!\![\boldsymbol{F}_1]_{m,:}\hat{ \boldsymbol{\omega} }_{n}\!\!-\!\nu_{p^{g}_{m,n}}\mu^g_{m,n},\label{nu_p_g_mn}
\end{align}
and $\hat{ \boldsymbol{\omega} }_{n}\triangleq [\hat{{\omega}}_{1,n},\cdots,\hat{{\omega} }_{M,n}]^T$, $\mu^g_{m,n}$ can be obtained by \eqref{v_s}, $\hat{{\omega}}_{m,n}$ and ${\nu}_{{\omega }_{m,n}}$ denote the available mean and variance, respectively, which are given in \eqref{v_gamma_nm}. 

Next, in the forward direction, based on the message $m_{f_{\boldsymbol{G}} \rightarrow{\tilde{g}}_{m,n}}(\tilde{g}_{m,n})$, we first apply UAMP in \emph{Part} $\uppercase\expandafter{\romannumeral3}$ where the intensive connections exists. Specifically, UAMP suggests that the message from function node $\omega_{m,n}$ to variable node $f_{\omega_{m,n}}$ is
\begin{align}
	m_{\omega_{m,n} \rightarrow f_{{\omega}_{m,n}}}(\omega_{m,n})\propto \mathcal{N}(\omega_{m,n};q^{g}_{m,n},\nu_{q^{g}_{m,n}}),
\end{align}
with mean $q^{g}_{m,n}$ and variance $\nu_{q^{g}_{m,n}}$, which are given by
\begin{align}\label{v_q_g_mn}
	{\nu}_{q^{g}_{m,n}}=1/ \langle\boldsymbol{\nu}_{\boldsymbol{\mu}^g_n}\rangle ,~~~~~
	{{q}^{g}_{m,n}}\!\!=\!\hat{\omega}_{m,n}\!\!+\!{\nu}_{q^{g}_{m,n}}[\boldsymbol{F}_1]_{m,:}{ \boldsymbol{\mu}^g_n},
\end{align}
where $\boldsymbol{\nu}_{\boldsymbol{\mu}^g_n}$ and ${\boldsymbol{\mu}^g_n}$ are respectively the vector-form of ${\nu}_{\mu^g_{m,n}}$ and $ \mu^g_{m,n}$, which is given by
\begin{align}
	{\nu}_{\mu^g_{m,n}}\!\!&=\!\!{1}/({\nu}_{p^{g}_{m,n}}\!\!+\!\rarrow{\nu}_{\tilde{g}_{m,n}}),~
	\mu^g_{m,n}\!\!=\!\!{\nu}_{\mu^g_{m,n}} (\rarrow{\nu}_{\tilde{g}_{m,n}}\!\!-\!{p}^{g}_{m,n}).\label{v_s}
\end{align}
Then, variational message passing is adopted to compute the message
\begin{align}\label{gamma_g_nm}
	& m_{f_{\omega_{m,n}}\rightarrow {{\gamma}^{g}_{m,n}} }\!({{\gamma}^{g}_{m,n}})  \nonumber \\
	& \propto \exp \left\{ \int  \mathfrak{b}(\omega_{m,n}) \log f_{\omega_{m,n}}(\omega_{m,n})d{\omega_{m,n}} \right\} ,
\end{align}
where $\mathfrak{b}( \omega_{m,n}) \propto  m_{{\omega_{m,n}} \rightarrow f_{\omega_{m,n}} }({\omega_{m,n}})  m_{f_{\omega_{m,n}}\rightarrow {\omega_{m,n}} }({\omega_{m,n}})$. As a multiplication of several Gaussian functions, $ \mathfrak{b}\left({\omega}_{m,n}\right)$ is still Gaussian, i.e., $ \mathfrak{b}\left({\omega}_{m,n}\right) \propto \mathcal{N} (\omega_{m,n};\hat{\omega}_{m,n},\nu_{\omega_{m,n}}) $ with  
\begin{align}\label{v_gamma_nm}
	\nu_{\omega_{m,n}} =\frac{{\nu}_{q^g_{m,n}}}{1+{\nu}_{q^{g}_{m,n}}\hat{\gamma}^{g}_{m,n}}, ~~~~
	\hat{\omega}_{m,n} =\frac{q^g_{m,n}}{1+{\nu}_{q^{g}_{m,n}}\hat{\gamma}^{g}_{m,n}}.  
\end{align}
Then, according to the above, we have 
\begin{align}
	m_{f_{\omega_{m,n}}\!\rightarrow {{\gamma}^{g}_{m,n}} }({{\gamma}^{g}_{m,n}}) \!\propto\! \exp \big\{ \!\!-\!\!\frac{{\gamma}^{g}_{m,n}}{2}(|\hat{\omega}_{m,n}|^2\!\!+\!\nu_{\omega_{m,n}} )  \big\}. 
\end{align}

As shown in the Part $\uppercase\expandafter{\romannumeral3}$ of Fig. \ref{fig:FactorGraph}, due to the symmetry between $\{\omega_{m,n}\}$ and $\{\sigma_{n,k}\}$, we can directly obtain the messages about $\{\sigma_{n,k}\}$ by replacing the corresponding parameters. 

The proposed algorithm is summarized in Algorithm \ref{algorithm}.
{The complexity of Part $\uppercase\expandafter{\romannumeral1}$ and $\uppercase\expandafter{\romannumeral2}$ is $\mathcal{O}(NLKM)$, and the complexity involved in Part $\uppercase\expandafter{\romannumeral3}$ is $\mathcal{O}(NK\log_{2}{N} )$ or $\mathcal{O}(NM\log_{2}{M})$ thanks to the FFT in \eqref{nu_p_g_mn} and \eqref{v_q_g_mn}. The complexity is lower than that of the channel estimation method in \cite{LiuJSAC2020}, which is $\mathcal{O}(N^2M^2)$.}
\begin{algorithm}\footnotesize
	\setstretch{0.8}
    \caption{\footnotesize{The proposed algorithm}}
	\textbf{Input:} $\boldsymbol{\Phi}$, $\zeta$ and the maximum number of iteration $I_{max}$.
	\\
	\textbf{Initialize:} $\hat{\omega}_{m,n}\!=\!0, \nu_{{\omega}_{m,n}}\!=\!1, {s}^g_{m,n}\!=\!0, \hat{\gamma}^{g}_{m,n}\!=\!1, \epsilon^{g}_{n}\!=\!0.001, \hat{\sigma}_{n,k}\!=\!0, \nu_{{\sigma}_{n,k}}\!=\!1,{s}^h_{n,k}\!=\!0, \hat{\gamma}^h_{n,k}\!=\!1, \epsilon^{h}_{k}\!=\!0.001$, $\forall m,k,n,j$. 
	\\
	\textbf{Repeat:}
	\begin{algorithmic}[1]
		\STATE $\forall j$: update $\boldsymbol{q}_{j}$ and $\boldsymbol{\nu}_{\boldsymbol{q}_{j}}$ as the Line 4 of the Algorithm 1 in \cite{GuoTWC2023}
		\STATE $\forall m,k,n$: update $\rarrow{\nu}_{g_{m,k,n}}$, $\rarrow{g}_{m,k,n}$ with (\ref{right_v_g_nmk}), (\ref{right_g_nmk});\\		
		\STATE $\forall m,n$: update $\rarrow{\nu}_{g_{m,n}}$, $\rarrow{g}_{m,n}$, $ \rarrow{\nu}_{\tilde{g}_{m,n}}$, $ \rarrow{\tilde{g}}_{m,n}$ with (\ref{right_v_g_nm}), (\ref{right_g_nm}), (\ref{right_v_z_nm});\\		
		\STATE $\forall m,n$: update ${\nu}_{\mu^g_{m,n}}, {\mu^g_{m,n}},{\nu}_{q^{g}_{m,n}}, {q^{g}_{m,n}}$, $\nu_{\omega_{m,n}}$, $\hat{\omega}_{m,n}$, $\hat{\gamma}^{g}_{m,n}$ with (\ref{v_s}), (\ref{v_q_g_mn}), (\ref{v_gamma_nm}), (\ref{gamma_hat});\\	
		\STATE $\forall n$: update ${\epsilon}^{g}_{n}$ with (\ref{epsilon_g});\\						
		\STATE $\forall m,n$: update ${\nu}_{p^g_{m,n}}, {p^g_{m,n}},{\nu}_{g^{\prime}_{m,n}}, {g^{\prime}_{m,n}}$, $\nu_{g_{m,n}}$, $\hat{g}_{m,n}$, $\larrow{\nu}_{g_{m,n}}$, $\larrow{g}_{m,n}$ with (\ref{nu_p_g_mn}), (\ref{tilde_g_mn}), (\ref{hat_g_mn}) and the Line 15 of Algorithm 1 in \cite{GuoTWC2023};\\	
		\STATE $\forall k,n$: {repeat step 2 to step 6 after replacing the corresponding parameters related to $\{\tilde{h}_{k,n}\}$ and $\{\sigma_{n,k}\}$;}
		\STATE $\forall j,n$: update $\larrow{\nu}_{{s}_{j,n}}$, $ \larrow{s}_{j,n}$ with \ref{le_s_jn} and \ref{s_jn}; \\
		\end{algorithmic}
\noindent \textbf{{Until}} $\frac{ \sum_{m,n} | \hat{\omega}_{m,n}-{\omega}_{m,n}|^2}{ \sum_{m,n} |{\omega}_{m,n}|^2}< \zeta$ and $\frac{ \sum_{k,n} | \hat{\sigma}_{n,k}-{\sigma}_{n,k}|^2}{\sum_{k,n} |{\sigma}_{n,k}|^2}< \zeta$ or the number of iteration is more than $I_{max}$.\\
\textbf{{Output:}} $ \hat{\omega}_{m,n}$, $ \hat{\sigma}_{n,k}$.
\label{algorithm}
\end{algorithm}

\section{Simulation Results}\label{sec:Simulation}

{In this section, we provide numerical experiments to demonstrate the superior performance of the proposed message passing algorithm. The most relevant work for comparison is \cite{LiuJSAC2020}, where a matrix-Calibration-based (MCB) channel estimation algorithm is proposed. In addition, we also include the performance of the oracle least-square (LS) estimator as a bound, which assumes the knowledge of the support of the sparse channels.} {It is noted that there exists an inevitable scaling ambiguity in the channel estimation, which is removed in the calculation of the normalized mean square error (NMSE)}\footnote{{The scaling ambiguity in channel matrix estimation does not hamper the downstream RIS beamforming design as shown in \cite{WangSPL2020,HeWCL2020}.}}. In our simulations, we set $M=K=N=32$, $N_1=4$ and $N_2=8$. The RIS phase matrix $\boldsymbol{\Phi}$ is {selected} as a partial DFT matrix. We assume a Rician channel comprising a line-of-sight path and a number of non line-of-sight paths and the Rician factor is set to 13.2dB \cite{WangSPL2020}. The number of paths for mmWave channels $\boldsymbol{G}$ and $\boldsymbol{h}_k$ are respectively set to $P=3$ and $P^{\prime}=3$, where the AoA and AoD parameters are uniformly generated from $[-\pi/2, \pi/2]$ and not necessarily lie on the discretized grid. The threshold $\zeta=\!10^{-3}$ and $I_{max}=30$. 
	
	\begin{figure}[t]\footnotesize
		\centering
		\includegraphics[width=0.9\columnwidth]{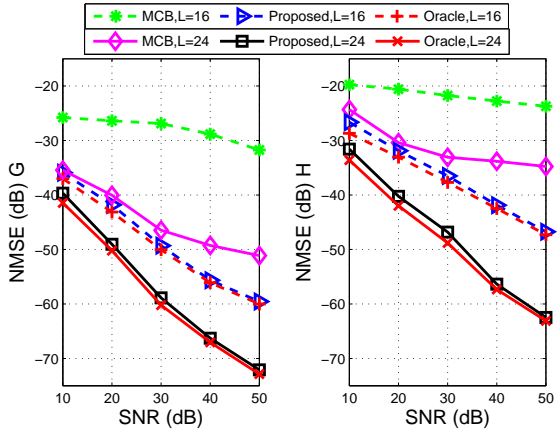}
		\caption{\footnotesize{ NMSE performance of the estimators versus SNR.}} \label{fig:L_16_24}
	\end{figure}
	
	\begin{figure}[t]\footnotesize
	\centering
	\includegraphics[width=0.9\columnwidth]{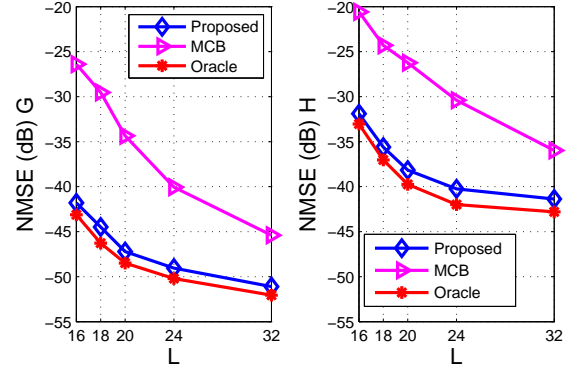}
	\caption{\footnotesize{ NMSE performance of the estimators versus SNR for different $L$.}} \label{fig:var_L}
\end{figure}
	
In Fig. \ref{fig:L_16_24}, we compare the NMSE performance versus SNR of the estimators with different values of $L$. {As shown by the results, the proposed method performs significantly better than the MCB method}, especially when $L$ is relatively small. It is noted that a smaller $L$ (the number of RIS phase configurations needed for channel estimation) is highly desirable to reduce the training overhead and latency. The results also show that the proposed method can achieve performance close to that of the oracle LS estimator for different $L$, even for small $L$. Next, we vary the value of $L$ and examine the performance of the estimators in Fig. \ref{fig:var_L}, where the SNR is set to 20dB. 
It can be seen that the performances of all estimators improve as expected with $L$. {To achieve the same MSE performance, the proposed algorithm uses significantly smaller $L$, indicating that the overhead and latency due to the channel estimation are greatly reduced. } %  about 42.2\% and 43.7\% overhead for NMSE=-45dB of $\boldsymbol{\Omega}$ and NMSE=-35dB of $\boldsymbol{\Sigma}$, respectively. }

{\bibliographystyle{IEEEtran}
\bibliography{IEEEabrv,bibliography}}

\begin{thebibliography}{10}
\providecommand{\url}[1]{#1}
\csname url@samestyle\endcsname
\providecommand{\newblock}{\relax}
\providecommand{\bibinfo}[2]{#2}
\providecommand{\BIBentrySTDinterwordspacing}{\spaceskip=0pt\relax}
\providecommand{\BIBentryALTinterwordstretchfactor}{4}
\providecommand{\BIBentryALTinterwordspacing}{\spaceskip=\fontdimen2\font plus
\BIBentryALTinterwordstretchfactor\fontdimen3\font minus
  \fontdimen4\font\relax}
\providecommand{\BIBforeignlanguage}[2]{{%
\expandafter\ifx\csname l@#1\endcsname\relax
\typeout{** WARNING: IEEEtran.bst: No hyphenation pattern has been}%
\typeout{** loaded for the language `#1'. Using the pattern for}%
\typeout{** the default language instead.}%
\else
\language=\csname l@#1\endcsname
\fi
#2}}
\providecommand{\BIBdecl}{\relax}
\BIBdecl

\bibitem{WangSPL2020}
P.~Wang, J.~Fang, H.~Duan, and H.~Li, ``Compressed channel estimation for
  intelligent reflecting surface-assisted millimeter wave systems,''
  \emph{{IEEE} Signal Processing Lett.}, vol.~27, pp. 905--909, 2020.

\bibitem{Huang2019TWC}
C.~Huang, A.~Zappone, G.~C. Alexandropoulos, M.~Debbah, and C.~Yuen,
  ``Reconfigurable intelligent surfaces for energy efficiency in wireless
  communication,'' \emph{{IEEE} Trans. Wireless Commun.}, vol.~18, no.~8, pp.
  4157--4170, 2019.

\bibitem{PengTCOM2022}
Z.~Peng, G.~Zhou, C.~Pan, H.~Ren, A.~L. Swindlehurst, P.~Popovski, and G.~Wu,
  ``Channel estimation for {RIS}-aided multi-user mm{W}ave systems with uniform
  planar arrays,'' \emph{{IEEE} Trans. Commun.}, vol.~70, no.~12, pp.
  8105--8122, 2022.

\bibitem{HeWCL2020}
Z.-Q. He and X.~Yuan, ``Cascaded channel estimation for large intelligent
  metasurface assisted massive {MIMO},'' \emph{{IEEE} Wireless Commun. Lett.},
  vol.~9, no.~2, pp. 210--214, 2020.

\bibitem{LiuJSAC2020}
H.~Liu, X.~Yuan, and Y.-J.~A. Zhang, ``Matrix-calibration-based cascaded
  channel estimation for reconfigurable intelligent surface assisted multiuser
  {MIMO},'' \emph{{IEEE} J. Sel. Areas Commun.}, vol.~38, no.~11, pp.
  2621--2636, 2020.

\bibitem{WangTWC2020}
Z.~Wang, L.~Liu, and S.~Cui, ``Channel estimation for intelligent reflecting
  surface assisted multiuser communications\: Framework, algorithms, and
  analysis,'' \emph{{IEEE} Trans. Wireless Commun.}, vol.~19, no.~10, pp.
  6607--6620, 2020.

\bibitem{ZhangCL2022}
X.~Zhang, X.~Shao, Y.~Guo, Y.~Lu, and L.~Cheng, ``Sparsity-structured
  tensor-aided channel estimation for {RIS-}assisted {MIMO} communications,''
  \emph{{IEEE} Commun. Lett.}, vol.~26, no.~10, pp. 2460--2464, 2022.

\bibitem{GuoTWC2023}
Y.~Guo, P.~Sun, Z.~Yuan, C.~Huang, Q.~Guo, Z.~Wang, and C.~Yuen, ``Efficient
  channel estimation for {RIS-}aided {MIMO} communications with unitary
  approximate message passing,'' \emph{{IEEE} Trans. Wireless Commun.},
  vol.~22, no.~2, pp. 1403--1416, 2023.

\bibitem{2015GuoApproximate}
\BIBentryALTinterwordspacing
Q.~Guo and J.~Xi, ``Approximate message passing with unitary transformation,''
  \emph{CoRR}, vol. abs/1504.04799, 2015. [Online]. Available:
  \url{http://arxiv.org/abs/1504.04799}
\BIBentrySTDinterwordspacing

\bibitem{YuanTSP2021}
Z.~Yuan, Q.~Guo, and M.~Luo, ``Approximate message passing with unitary
  transformation for robust bilinear recovery,'' \emph{{IEEE} Trans. Signal
  Processing}, vol.~69, pp. 617--630, 2021.

\bibitem{UAMPSBL}
M.~Luo, Q.~Guo, M.~Jin, Y.~C. Eldar, D.~Huang, and X.~Meng, ``Unitary
  approximate message passing for sparse bayesian learning,'' \emph{IEEE Trans.
  Signal Process.}, vol.~69, pp. 6023--6039, 2021.

\end{thebibliography}

\end{document}